\documentclass{pasj00}

\draft

\begin{document}
\SetRunningHead{S. Kato}{A Vertical Resonance Model of QPOs}
\Received{2005/04/20}
\Accepted{2005/06/13}

\title{A Vertical Resonance of G-Mode Oscillations in Warped Disks
and QPOs in Low-Mass X-Ray Binaries}

\author{Shoji \textsc{Kato}}
\affil{Department of Informatics, Nara Sangyo University, Ikoma-gun,
       Nara 636-8503}
\email{kato@io.nara-su.ac.jp, kato@kusastro.kyoto-u.ac.jp}


%


\maketitle

\begin{abstract}

Resonant oscillations in warped disks are examined in order to explain
high-frequency QPOs and horizontal-branch QPOs in low-mass X-ray
binaries.
Different from our previous work, addressed to the same subjects,
we relax in this paper the assumption that the disks are isothermal in
the vertical direction.
That is, the pressure, density, and temperature are assumed to be 
distributed in the vertical direction with a polytropic relation, and
the polytropic index changes as the disk state changes.
By this generalization and by some subsidiaries we can qualitatively explain, 
within the framework
of our resonance model, observed large frequency variations  
in neutron-star QPOs and little variations in black-hole QPOs.
We consider vertical
resonances of g-mode oscillations, since they are most appropriate 
to explain observations.

\end{abstract}

\section{Introduction}

In this paper we are interested in possible origins of high-frequency 
quasi-periodic oscillations (QPOs)
in neutron-star and black-hole low-mass X-ray binaries (LMXBs), 
and additionally
in horizontal-branch QPOs in luminous neutron-star LMXBs (Z sources).
One of the important characteristics of kHz QPOs in neutron-star LMXBs
is that they usually occur in a pair, and their frequencies change 
with time (see, e.g., van der Klis 2000, 2004 for reviews).
The time change occurs in such a way as their frequency ratio decreases
with an increase of frequencies.
In the case of black-hole LMXBs, high-frequency QPOs usually also occur
in a pair, but their frequencies change little with time, 
the ratio being kept close to 2 : 3 (see van der Klis 2004 for a review).
In Z sources, another important characteristic of QPOs is present.
It is a strong correlation between kHz QPOs and horizontal branch QPOs
in their frequencies (Psaltis et al. 1999).
That is,  the frequency of the lower 
kHz QPO, $\nu_{\rm L}$, and that of the horizontal branch QPOs,  
$\nu_{\rm HBO}$, are correlated
in each object so that $\nu_{\rm HBO}\sim 0.08
\nu_{\rm L}$.

Paying attention to the closeness to 2 : 3 of the frequency ratio in 
high-frequency QPOs in balck hole LMXBs, Abramowicz and Klu{\' z}niak (2001)
and Klu{\' z}niak and Abramowicz (2001) presented a resonance model of
disk oscillations to explain the QPOs.
On the other hand, the importance of an external deformation of disks on 
resonance processes
has been emphasized by Kato (2003, 2004a, 2004b), Klu{\' z}niak et
al. (2004), and Lee, Abramowicz, and Klu{\' z}niak (2004).
Kato (2003, 2004a, 2004b) especially considered models of resonant 
oscillations on warped disks.
The models were extended to a case where the warp had precession
(Kato 2005).
In these warp models, the temperature is assumed to always 
distribute isothermally in the vertical direction.
In real situations, however, this will not be the case.
The vertical structure of disks will change by the 
difference of the disk state resulting from difference of the 
mass-accretion rate.
Considering this, we assume in this paper that the pressure, density,
and temperature distributions in the vertical direction are polytropic, and 
that the polytropic index changes with time.
By adopting this model, we can easily obtain the time variations
of the resonant oscillations as a result of changes of the polytropic 
index.\footnote{In Kato (2005), time variations of QPOs was interpreted 
as a result of time change of precession of warps.} 
The purpose of this paper is to demonstrate this generalization and 
to apply the results to QPOs.

In warped disks there are four types of resonant oscillations due to the 
possible four combinations of two types of oscillations (g-mode and
p-mode oscillations)
and two types of resonances (vertical and horizontal resonances).
Without considering all cases, we consider here only a case where  
the time variations of QPOs can be well explained.
This is the case where disk oscillations are g-modes, 
and the resonances occur through vertical 
motions.\footnote{ In Kato (2005), horizontal resonances of g-mode oscillations
were considered.}

\section{Vertical Resonances of G-Mode Oscillations in Warped Disks}

Details of the non-linear resonant oscillations on warped disks are
presented by Kato (2003, 2004a, 2004b).
An overview of the model is sketched in figure 1 
of Kato (2004a).

We consider geometrically thin relativistic disks rotating with 
angular velocity, $\Omega(r)$.
The epicyclic frequency on the disk is denoted by $\kappa(r)$.
The oscillations on  geometrically thin disks are classified into g-mode
and p-mode oscillations (see, e.g., Kato et al. 1998; Kato 2001).
In simplified disks the oscillations are further sub-classified by the 
set ($m$, $n$), where
$m=$(0, 1, 2...) is the number of nodes in the azimuthal direction, and 
$n=$(0, 1, 2...) is the number related to nodes in the vertical direction.
That is, $n$ represents the number of nodes that $u_r$ (the radial 
component of velocity associated with oscillations) has in
the vertical direction.
It is noted, however, that $u_z$ (the vertical component of velocity
associated with oscillations) has ($n-1$) nodes in the vertical direction.
In the case of $n=0$,  $u_z=0$ and 
a series of g-modes start from $n=1$.
    
A warp is generally a global pattern on disks with $m=n=1$.
The warp may have precession, but here we assume that it has no
precession in order to consider idealized cases.
On a disk deformed by the warp, we superpose g-mode oscillations with
arbitrary $m$ and $n$ ($n$ is taken to be 1 or 2, later).
A g-mode oscillation with the frequency $\omega$ and ($m$, $n$) has 
a relatively large amplitude, global pattern only around the radius where
\begin{equation}
    (\omega-m\Omega)^2-\kappa^2 = 0
\label{1.1}
\end{equation}
is satisfied.
This can be understood if the dispersion 
relation for local perturbations is considered. 
That is, the dispersion relation shows that the region of 
$(\omega-m\Omega)^2-\kappa^2 >0$ is an
evanescent region of the g-mode oscillations.
In the region where 
$(\omega-m\Omega)^2$ is smaller than $\kappa^2$, on the other hand,
the oscillations have very short wavelengths in the radial direction 
when geometrically thin disks are considered.   

A non-linear interaction of a g-mode oscillation with the warp
produces an oscillation with $\omega$, ${\tilde m}$, 
and ${\tilde n}$, where ${\tilde m}=m\pm 1$ and ${\tilde n}=n\pm 1$
(these oscillations are called hereafter intermediate oscillations).
These intermediate oscillations resonantly interact with the disk at 
the radius where 
the dispersion relation for these intermediate oscillations is 
satisfied [see Kato (2004b) for detailed discussions].
There are two types of resonances, corresponding two types of wave modes.
One is resonances that occur through motions in the vertical direction
(vertical resonances);
the other is those through motions in the radial direction
(horizontal resonances)(see, e.g., Kato 2004a,b).
In this paper we are interested in vertical resonances.
They occur 
around the radius where 
\begin{equation}
   (\omega-{\tilde m}\Omega)^2-\Psi_{\tilde n} \Omega^2 \sim 0
\label{1.2}
\end{equation}
is satisfied (see e.g., Kato 2004a, b, and the Appendix in this paper), where 
$\Psi_{\tilde n}$ is a number related to
${\tilde n}$ and $N$ (a polytropic index specifying the vertical
structure of disks, see the Appendix).
Hereafter, we are particularly interested in cases where ${\tilde n}=2$ 
(assuming $n=1$) and ${\tilde n}=3$ (assuming $n=2$).
The quantities $\Psi_{\tilde n}$ in these cases are
\begin{equation}
    \Psi_2=2+{1\over N} \quad {\rm and}\quad  \Psi_3 =3+{3\over N},
\label{1.3}
\end{equation}
as shown in the Appendix.
In the limit where the disk is isothermal in the vertical direction, 
$N=\infty$ and we have $\Psi_2=2$ and $\Psi_3=3$.
They are cases that have been considered in previous papers (Kato 2003,
2004a, 2004b).

Combining equations (\ref{1.1}) and (\ref{1.2}), we find that the resonances 
occur at radii of 
\begin{equation}
    \kappa=(\Psi_2^{1/2}-1)\Omega \quad {\rm and} \quad
     \kappa=(\Psi_3^{1/2}-1)\Omega,
\label{1.4}
\end{equation}
for ${\tilde n}=2$ and ${\tilde n}=3$, respectively.
After this resonance the intermediate oscillations feedback to the
original oscillations by non-linear interaction with the original 
oscillations themselves.
The results are amplification or dampening of the original oscillations,
depending on the types of oscillations and resonances (Kato 2004b).

Detailed examination shows that when ${\tilde m}=m-1$,
the frequencies of the g-mode oscillations that resonantly
interact with the disk at $\kappa=(\Psi_{\tilde n}^{1/2}-1)\Omega$ are 
$m\Omega-\kappa$ at the radius.
On the other hand, when ${\tilde m}=m+1$ the frequencies of the g-mode 
oscillations that resonantly interact with the
disk at $\kappa=(\Psi_{\tilde n}^{1/2}-1)\Omega$ are 
$m\Omega+\kappa$ there  (e.g., Kato 2004a).
The most observable oscillations are those with small $m$'s;
otherwise, they are phase-mixed.
Hence, the frequencies, $\omega$, of 
typical non-axisymmetric resonant oscillations are
$\Omega-\kappa$ (i.e., $m=1, {\tilde m}=0)$, $\Omega+\kappa$
(i.e., $m=1, {\tilde m}=2)$, and $2\Omega-\kappa$ 
(i.e., $m=2, {\tilde m}=1)$.
As axisymmetric resonant oscillations we have $\kappa$  
(i.e., $m=0, {\tilde m}=1)$.
Axisymmetric oscillations, however, will be less observable.
Hence, we think that the above non-axisymmetric oscillations 
are related to the horizontal branch QPOs, upper and lower kHz QPOs. 
Considering this, we introduce
\begin{equation}
  \omega_{\rm H}=\Omega+\kappa, \quad \omega_{\rm L}=2\Omega-\kappa, \quad
  {\rm and} \quad \omega_{\rm HBO}=\Omega-\kappa,
\label{1.5}    
\end{equation}
and examine their frequencies at the resonance radius.

\section{Frequencies and Their Variations of Resonant Oscillations}

In the limit of isothermal disks in the vertical direction, i.e.,
$N=\infty$, the resonance occurs at $\kappa=(\sqrt{2}-1)\Omega$
when ${\tilde n}=2$ and at $\kappa=(\sqrt{3}-1)\Omega$ when
${\tilde n}=3$.
These radii are $3.62r_g$ and $6.46r_g$, respectively (e.g., Kato 2004a).
The temperature distribution in the vertical direction in disks is, however, 
generally not isothermal.
Furthermore, it changes with time, depending on the state of the disks.
In figure 1, the $r/r_g$--$\Psi$ relation given by
$\kappa=(\Psi^{1/2}-1)\Omega$ is shown for the range of
$\Psi=2.0$--$4.0$.
In the case of ${\tilde n}=2$, $\Psi$ (i.e., $\Psi_2$) changes from 2 
(for $\gamma=1$ or $N=\infty$)
to 2.67 ($\gamma=5/3$ or $N=1.5$), while $\Psi_3$ changes from 3
($\gamma=1$) to 5 (for $\gamma=5/3$) in the case of ${\tilde n}=3$.
The ranges of variations of $\Psi_2$ and $\Psi_3$ are also shown in
figure 1.

If the value of $\Psi$ is specified, the resonant radius is obtained.
Then, from equation (\ref{1.5}), the frequencies of the resonant oscillations, 
$\omega_{\rm H}$, $\omega_{\rm L}$, and 
$\omega_{\rm HBO}$ are derived as functions of $\Psi$.
The relations among these frequencies are shown in figure 2 as
functions of $\omega_{\rm H}$.
The relation between $2\omega_{\rm HBO}$ and $\omega_{\rm H}$ is
also shown.
Important points are that: i) figure 2 is free from the mass of
the central source, if the both horizontal and vertical scales are
normalized by $(M/M_\odot)^{-1}$, and
ii) the curves in figure 2 are universal as long as g-mode oscillations
are concerned.
They are free from $N$ (and $\Psi$), and even free from the presence or
absence of precession (see Kato 2005).
That is, the frequencies of the resonant oscillations change along
the curves when $N$ (and thus $\Psi$) is changed.
The relation between $\omega_{\rm H}$ and $\Psi$ is shown in figure 3.
In general, an increase of $\Psi$ decreases $\omega_{\rm H}$.
The range of the variation of $\omega_{\rm H}$ by a change of $\Psi$ is also 
shown in figure 2.

\section{Summary and Numerical Estimate}

The main results of this paper are summarized in figure 2.
The figure should be compared with figure 2.6 of van der Klis (2004),
which summarizes observational data concerning QPO frequencies
on a frequency--frequency diagram.
This comparison suggests that the resonant oscillations specified by
$\omega_{\rm H}$, $\omega_{\rm L}$, and $\omega_{\rm HBO}$ well
correspond to the upper and lower kHz QPOs and the horizontal branch QPOs
in neutron-star LMXBs, respectively.
Adopting these identification, we qualitatively explain 
some basic observational characteristics of the QPOs.

The kHz QPOs in neutron-star LMXBs usually appear in a pair, and
the separation frequency of the twin peaks decreases as the peak frequencies
increase.
This observational characteristic is derived in our model, as
shown in figure 2 (see the curves of $\omega_{\rm L}$ and $\omega_{\rm H}$).
In our model the frequency change of QPOs is the result of a change of 
the temperature distribution in the vertical direction.
As the temperature distribution in the vertical direction approaches to
an isothermal one [i.e., $\Psi_2$ (the case of ${\tilde n}=2$) tends to
2.0 or $\Psi_3$ (the case of ${\tilde n}=3$) tends to 3.0],
the resonance radius becomes smaller (see figure 1) and
the frequencies of resonant oscillations increase.
Observations show that the frequencies of the pair QPOs increase 
with an increase of the
mass-accretion rate (van der Klis et al. 1997).
Hence, if our model is correct, it suggests that the temperature
distribution in the vertical direction approaches to the isothermal one
as the mass-accretion rate increases.
Here, let us make a quantitative estimate.
Figure 2 shows that the 3 : 2 ratio of $\omega_{\rm H}$ and 
$\omega_{\rm L}$ is realized when 
log$[\omega_{\rm H}(M/M_\odot)]\sim 2.93$,
i.e., $\omega_{\rm H}=851(M/M_\odot)^{-1}$.
This occurs for $\Psi_3=3.24$ (i.e., ${\tilde n}=3$), as shown in figure 3.
This gives $1/N=0.08$ and $\gamma=1.08$.
The resonance radius is then at $8.33r_g$, as shown in figure 1.

One of another prominent correlations among QPO frequencies in
neutron-star LMXBs is that the frequencies of the horizontal-branch QPOs,
$\nu_{\rm HBO}$,
are correlated with the frequencies of lower kHz QPOs, $\nu_{\rm L}$,
by $\nu_{\rm HBO}\sim 0.08\nu_{\rm L}$.
In order to compare our rersults with the observations,
the curve of the $0.08\omega_{\rm L}$--$\omega_{\rm H}$ 
relation is drawn in figure 2.
The curve crosses the curve labelled by $\omega_{\rm HBO}$ around
log$[\omega_{\rm H}(M/M_\odot)]\sim 2.46$,
i.e., $\omega_{\rm H}=288(M/M_\odot)^{-1}$.
Figure 3 then shows $\Psi_3\sim 3.66$, which means
$3/N\sim 0.66$ or $\gamma\sim 1.22$.
This implies that the observed correlation can be explained if the
temperature distribution in the vertical direction is, on average, 
around $\gamma\sim 1.2$.
From figure 1 we see that the resonance occurs around
$r=18.0r_g$ when $\Psi_3=3.66$.
As shown above, both $\omega_{\rm H}$ : $\omega_{\rm L} =$ 3 : 2 and
$\omega_{\rm HBO}=0.08\omega_{\rm L}$ cannot be simultaneously
satisfied in a rigorous sense.
A slight larger coefficient than 0.08, say 0.09, is compatible with 
the ratio 3 : 2.

In the case of black-hole LMXBs, observations show that the 
frequencies of pair QPOs change little, and their ratio is kept to be 
close to 3 : 2, unlike the case of neutron-star kHz QPOs.
As discussed in the case of neutron-star QPOs, one possibility 
of explaining the
observed 3 : 2 is that it represents $\omega_{\rm H}$ :
$\omega_{\rm L}$.
Another one, which is better, is to consider the resonances of
${\tilde n}=2$ and to regard the ratio as $\omega_{\rm L}$ :
$2\omega_{\rm HBO}$.
As discussed in Kato (2004b), if we consider the resonance at $4.0r_g$,
$\omega_{\rm H}$ is equal to $\omega_{\rm L}$ and the ratio
$\omega_{\rm L}$(or $\omega_{\rm H}$) : $2\omega_{\rm HBO}$ is just
3 : 2.
Figure 1 shows that the resonance at $4.0r_g$ is realized 
for ${\tilde n}=2$ when $\Psi_2=2.25$,
which means $1/N=0.25$ (i.e., $\gamma=4/3$) in the present model.
Figure 2 (see also figure 3) shows that this occurs when 
log$[\omega_{\rm H}(M/M_\odot)]=3.33$, i.e., $\omega_{\rm H}=
2.14\times 10^3(M/M_\odot)^{-1}$.
It is noted that in the case where the resonance occurs through 
intermediate oscillations
of ${\tilde n}=2$ (not ${\tilde n}=3$), the frequency variation for a
change of $N$ (or $\gamma$) is weak (see the variation range of $\Psi_2$
in figure 2).
Furthermore, $\omega_{\rm H}=2.14\times 10^3(M/M_\odot)^{-1}$
is compatible with observed results derived by McClintock and
Remillard (2003).
It is, however, unclear why the balck-hole QPOs are 
the resonances of ${\tilde n}=2$ $(n=1)$, while the neutron-star 
QPOs are the resonances of ${\tilde n}=3$ $(n=2)$.

More quantitative comparisons of our model with observations would be
premature at the present stage, since the frequencies of vertical 
resonances are sensitive to the vertical structure of the disks
(the vertical structure is not always polytropic).
In other words, if our present model is correct, a comparison of our results
with observations would give good information concerning the vertical 
structure of disks.

\section{Discussion}

The present resonance model naturally explains some basic characteristics 
of the kHz QPOs of neutron-star LMXBs and the high-frequency QPOs of
black-hole LMXBs.
It is especially noted that the observed frequency change of kHz QPOs,
which is observationally related to a change of the mass-accretion rate, is
naturally explained if the vertical structure of disks changes with
the change of the mass-accretion rate.
In this sense, the present model seems to be superior to a precession
model of warps (Kato 2005).
In the latter model, a time variation of precession is required to
explain the time variation of the observed kHz QPO frequencies, and 
it is unclear whether such a variation of precession is generally expected 
theoretically.
The latter model with precession, however, can naturally explain the
observed hectohertz QPOs as a manifestation of the precession of warps.
In the present model of the vertical resonances, on the other hand, 
there is a problem concerning excitation.
The vertical resonances do not excite the g-mode
oscillations, but rather dampen them in the limit of $N=\infty$
(Kato 2004b).
We should carefully study in the future whether our previous results concerning
the stability of resonancs are correct and relevant, even when $N\not=\infty$.
Comparisons of the characteristics of the present model with those of the
precession model are given in table 1.

Many QPO models have been proposed so far.
Some of them connect the observed QPO frequencies with the characteristic
frequencies of disks, such as the orbital frequency, radial and vertical
epicyclic frequencies and others, or their combinations.
For example, in their precession model, Stella and Vietri (1998) 
identify the periastron precession frequency, $\Omega-\kappa$
[see the third equation in (\ref{1.5})], with the lower frequency of 
the kHz QPO.
An important issue in such models is where preferred radii to produce
specific frequencies exist.
Concerning this point, these models are classified into various types,
e.g., precession model (Stella, Vietri 1998), resonance model 
(Klu{\' z}niak, Abramowicz 2001; Abramowicz, Klu{\' z}niak 2001), and
beat-frequency model (Miller et al. 1998).
Our warp model is based on hydrodynamical wave phenomena and different from
the models mentioned above.
However, if we discuss the present model in connection with them, a warp 
can be regarded as a process to select preferred radii.

Finally, the applicability of the present model to cataclysmic variables
(CVs) is briefly discussed.
Mauche (2002) pointed out that the frequency correlation $\nu_{\rm
HBO}\sim 0.08 \nu_{\rm L}$ can be extended to CVs.
This has recently been confirmed by Warner and Woudt (2004).
That is, the DNO (dwarf novae oscillation)--QPO relation in
CVs is on the line of extension of the HBO--kHz QPO relation in X-ray
stars.
Furthermore, it is known that the DNOs in CVs change their frequencies,
accompaning harmonics (Warner, Woudt 2004).

The disks of CVs are Keplerian and $\Omega$ and $\kappa$ are almost
equal.
Hence, at a glance, the resonance conditions, i.e., equations (\ref{1.4}),
seem not to be realized anymore.
This is, however, not the case.
The observations show that DNOs usually occur in the phase of outbursts.
Near the transition front of outbursts, the disk thickness changes
sharply.
In such region, the derivation
of the resonance condition, $\kappa=(\Psi_3^{1/2}-1)\Omega$, will be
inaccurate, since the assumption of slow radial change of $H/r$ is
involved in the derivation.
If we assume, however, that the resonance condition,  
$\kappa=(\Psi_3^{1/2}-1)\Omega$,
is still valid, the resonances occur even in the Newtonian Kepler disks, if
$\Psi_3\sim 4$.
Since $\Psi_3=3+3/N$, $\Psi_3=4$ is realized when $1/N=1/3$, i.e.,
$\gamma=4/3$.
(As mentioned before, $\Psi_3=3$ for $\gamma=1$ and $\Psi_3=5$ for
$\gamma=5/3$.)
A vertical disk structure with $N=3$ will not be unrealistic.
This consideration suggests that the harmonic structure of DNOs 
can be interpreted as the oscillations of $2\Omega+\kappa$,
$\Omega+\kappa$, and $\kappa$, at the resonant radius.
Their frequency ratios are 3 : 2 : 1 in the Newtonian Kepler disks.
The observed frequency changes of DNOs are interpreted as being the result 
of a change of the
resonance radius by a change of the disk structure.
[See also Klu{\'z}niak et al. (2005) for an interpretation of DNOs.]  
The observed correlation between DNOs and QPOs in CVs, however, cannot be
explained by the present model.
Further considerations are needed.

\begin{longtable}{ccc}
  \caption{Comparison of two resonant models of g-mode oscillations.}
  \label{tab:LTsample}
  \hline\hline
       & Horizontal resonances (Kato 2005) & Vertical resonances 
                                 (present paper) \\
  \hline
\endhead
  \hline
\endfoot
  \hline
\endlastfoot
  Time variation & change of precession & change of vertical disk structure \\
  Hectohertz QPOs & precession &  precession ? \\
  Excitation &     $\bigcirc$(probably)          &  ?  \\
\end{longtable}

\appendix
\section{Vertical Oscillations of Polytropic Disks}

We employ cylindrical coordinates ($r$, $\varphi$, $z$), whose origin is 
at the center of the central object and the $z$-axis is in the vertical
direction perpendicular to the disk plane.
Let us consider a disk where the pressure, $p$, and density, $\rho$,
are related in the vertical direction with a polytropic relation, i.e.,
$p=K\rho^{1+1/N}$, where $N$ is the polytropic index.
Then, the vertical integration of the hydrostatic balance in the vertical
direction gives 
\begin{equation}
    T_0=T_{00}(r)\biggr(1-{z^2\over H^2}\biggr),
\label{A.1}
\end{equation}
\begin{equation}
    \rho_0=\rho_{00}(r)\biggr(1-{z^2\over H^2}\biggr)^N,
\label{A.2}
\end{equation}
and
\begin{equation}
    p_0=p_{00}(r)\biggr(1-{z^2\over H^2}\biggr)^{N+1},
\label{A.3}
\end{equation}
where subscript 0 represents the quantities in the equilibrium
state, and 00 represents those on the equatorial plane (e.g.,
Kato et al. 1998).
The half-thickness of the disk, $H(r)$, is related to $p_{00}$
and $\rho_{00}$ by
\begin{equation}
   \Omega_{\rm K}^2H^2=2(N+1){p_{00}\over \rho_{00}}.
\label{A.4}
\end{equation}

We consider here a vertical oscillation of the disk, 
assuming, for simplicity,  that it has no velocity in the horizontal 
direction.
The perturbation associated with the oscillation is assumed to be 
proportional to exp[$i(\omega t-m\varphi)$], where $\omega$ is the
frequency of the oscillation and $m$ is the number of arms in the 
azimuthal direction. 
If the oscillation occurs under the polytropic relation given 
above, the vertical component of the equation of motion gives
\begin{equation}
   i(\omega-m\Omega)u_z=-{\partial h_1\over\partial z},
\label{A.5}
\end{equation}
where $h_1=c_{\rm s}^2\rho_1/\rho_0$ and $u_z$ and $\rho_1$ are
the vertical velocity and the density perturbation associated
with the oscillation, respectively, and 
$c_{\rm s}^2=\gamma p_0/\rho_0$ with $\gamma=1+1/N$.
The equation of continuity is written as
\begin{equation}
     i(\omega-m\Omega)\rho_1+{\partial\over\partial z}(\rho_0 u_z)=0.
\label{A.6}
\end{equation}
  
Elimination of $u_z$ from the above two equations gives an 
equation for $h_1$ as
\begin{equation}
   {1\over\rho_0}{\partial\over\partial z}\biggr(\rho_0{\partial h_1
        \over  \partial z}\biggr)
     +{(\omega-m\Omega)^2\over c_{\rm s}^2}h_1=0,
\label{A.7}
\end{equation}
which is reduced to
\begin{equation}
    {\partial^2 h_1\over \partial x^2}-{2Nx\over 1-x^2}
            {\partial h_1\over\partial x}
     +{2N\over 1-x^2}{(\omega-m\Omega)^2 \over \Omega_{\rm K}^2}h_1=0
\label{A.8}
\end{equation}
by changing the independent variable from $z$ to a dimensionless 
variable $x$ defined by $x=z/H$.
This equation should be solved with a boundary condition that the
Lagrangian change of pressure is zero at the disk surface to
obtain eigen-values $\omega$'s.

The solutions of equation (\ref{A.8}) are easily obtained by expressing
$h_1$ in terms of a finite power series of $x$.
In the fundamental mode of oscillations, $h_1\propto x$ and
$(\omega-m\Omega)^2=\Omega^2$.
In the first overtone, $h_1\propto 1-(1+2N)x^2$ and $(\omega-m\Omega)^2
=(2+1/N)\Omega^2$.
In the second overtone we have $h_1\propto x-(1+2N/3)x^3$ and
$(\omega-m\Omega)^2=(3+3/N)\Omega^2$.
In this way we obtain the eigen-frequencies of vertical oscillations as
\begin{equation}
   (\omega-m\Omega)^2=\Psi_n\Omega^2,
\label{A.9}
\end{equation}
where for the fundamental ($n=1$), first overtone ($n=2$), and second
overtone ($n=3$) we have
\begin{equation}
    \Psi_1 =1,
\label{A.10}
\end{equation}
\begin{equation}
    \Psi_2 =2+{1\over N},
\label{A.11}
\end{equation}
and
\begin{equation}
    \Psi_3 =3+{3\over N}.
\label{A.12}
\end{equation}
For $\gamma=5/3$ and $\gamma=4/3$, we have $1/N=2/3$ and $1/3$,
respectively.
In the case of isothermal disks, $1/N=0$.

\bigskip
\leftskip=20pt
\parindent=-20pt
\par
{\bf References}
\par
Abramowicz, M. A., \& Klu{\' z}niak, W. 2001, A\&A, 374, L19 \par
Kato, S. 2001, PASJ, 53, 1\par 
Kato, S. 2003, PASJ, 55, 801\par
Kato, S. 2004a, PASJ, 56, 559 \par
Kato, S. 2004b, PASJ, 56, 905\par
Kato, S. 2005, PASJ, 57, L17 \par
Kato, S., Fukue, J., \& Mineshige, S. 1998, Black-Hole Accretion Disks 
  (Kyoto: Kyoto University Press)\par
Kluz{\' n}iak, W., \& Abramowicz, M. 2001, Acta Phys. Pol. B32, 3605   \par
Klu{\' z}niak, W., Abramowicz, M. A., Kato, S., Lee, W. H., \& Stergioulas,
   N. 2004, ApJ, 603, L89 \par 
Kluz{\' n}iak, W., Lasota, J-P., \& Abramowicz, M.A. 2005, astro-ph 0503151\par
Lee, W.H., Abramowicz, M.A., \& Klu{\' z}niak, W. 2004, ApJ, 603, L93 \par
Mauche, C. W. 2002, ApJ, 580, 423\par   
McClintock, J.E., \& Remillard, R.A. 2003, astro-ph/0306213 \par
Miller, M.C., Lamb, F.K., \& Psaltis, D. 1998, ApJ, 508,791\par
Psaltis, D., Belloni, T., \& van der Klis, M. 1999, ApJ, 520, 262\par
Stella, L., \& Vietri, M. 1998, ApJ, 492, L59 \par
van der Klis, M. 2000, ARA\&A, 38, 717    \par
van der Klis, M. 2004, astro-ph/0410551    \par
van der Klis, M., Wijnands, R. A. D., Horne, K., \& Chen, W. 1997, ApJ,
     481, L97 \par
Warner, B., \& Woudt, P. 2004, astro-ph/0409287, in ASP Conf. Ser. 
   The Astrophysics of Cataclysmic Variables
   and Related Objects, ed. J. M. Hameury \& P.-J. Lasota
    \par

\bigskip\bigskip

\begin{figure}
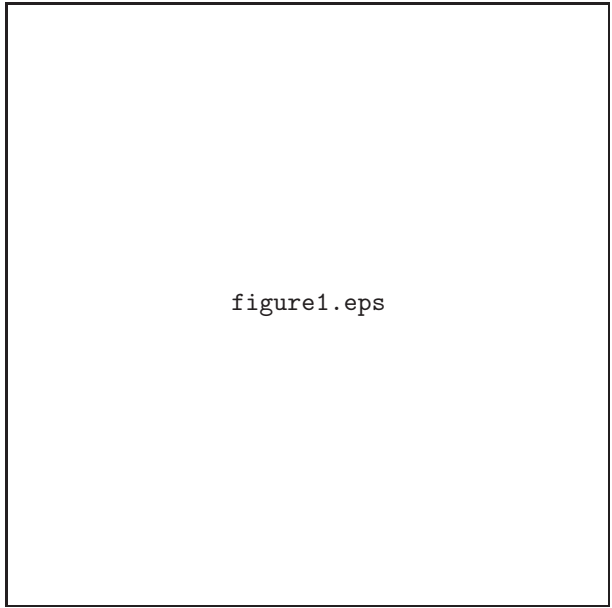

  \begin{center}
    \FigureFile(80mm,80mm){figure1.eps}
  \end{center}
  \caption{
The $r/r_g$--$\Psi$ relation obtained by solving  
the resonance condition $\kappa=(\Psi^{1/2}-1)\Omega$.
The disk is Keplerian with the Schwarzschild metric.
The ranges of variation of $\Psi_2$ and $\Psi_3$ are shown.}
\label{fig:figure 1}
\end{figure}

\begin{figure}
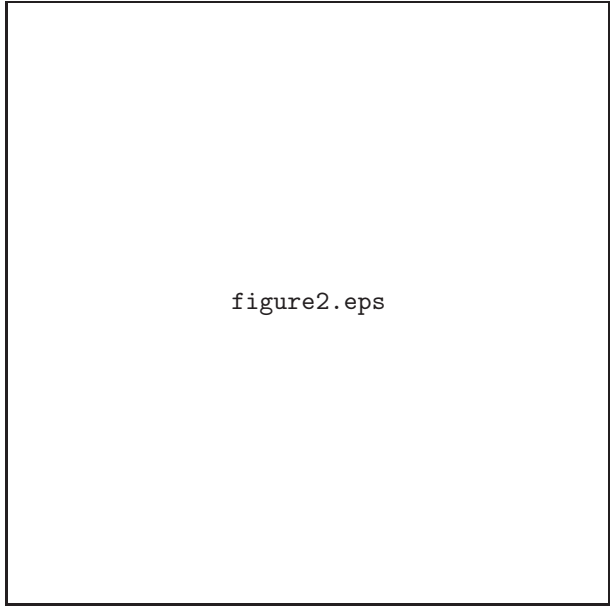

  \begin{center}
    \FigureFile(80mm,80mm){figure2.eps}
  \end{center}
  \caption{
Frequencies ($\omega_{\rm H}$, $\omega_{\rm L}$, $\omega_{\rm HBO}$,
and $2\omega_{\rm HBO}$)
as functions of $\omega_{\rm H}$.
For a comparison, the $0.08\omega_{\rm L}$--$\omega_{\rm H}$ relation
is also shown.
The ranges of variation of $\Psi_2$ and $\Psi_3$ are shown near the 
horizontal scale.} 
\label{fig:figure 2}
\end{figure}

\begin{figure}
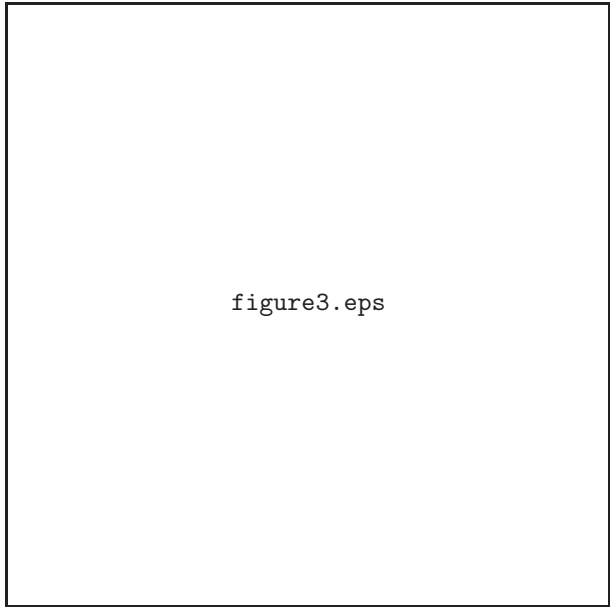

  \begin{center}
    \FigureFile(80mm,80mm){figure3.eps}
  \end{center}
  \caption{
$\Psi$--$\omega_{\rm H}$ relation. 
The disk is Keplerian with the Schwarzschild metric.
The ranges of variations of $\Psi_2$ and $\Psi_3$ are also shown.}
\label{fig:figure 3}
\end{figure}

\end{document}